\documentclass[twocolumn,showpacs,preprintnumbers,pra,amsmath,amssymb]{revtex4}

\usepackage{graphicx}% Include figure files
\usepackage{bm}% bold math
\begin{document}
\title{Reference-Frame-Independent  Measurement-Device-Independent Quantum Key Distribution with Uncharacterized Qubits}

\author{Won-Young Hwang}

\affiliation{Department of Physics Education, Chonnam National University, Gwangju 61186, Republic of Korea
}
%%%%%%%%%%%%%%%%%%%%%%%%%%%%%%%%%%%%%%%
\begin{abstract}
We propose a reference-frame-independent measurement-device-independent quantum key distribution with uncharacterized quantum bits. We show the security of the protocol. The protocol can also be useful for a channel that has a very low bit error rate but suffers a large, uncontrolled (but slow), unitary rotation.
\end{abstract}
%%%%%%%%%%%%%%%%%%%%%%%%%%%%%%%%%%%%%%%
\maketitle
%%%%%%%%%%%%%%%%%%%%%%%%%%%%%%%%%%%%%%%
\section{Introduction}
%%%%%%%%%%%%%%%%%%%%%%%%%%%%%%%%%%%%%%%
The quantum key distribution (QKD) is an appealing field theoretically as well as practically \cite{Sca09}. After the security of the QKD had been shown for ideal devices \cite{May01,Sho00}, problems due to imperfect devices surfaced. Although the problems due to imperfect source were resolved \cite{Hwa03}, those due to imperfect detectors still remained \cite{Mak06, Qi07, Fun07}.
Device-independent QKD's elegantly overcome the problems \cite{Aci07}. However, a device-independent QKD is not yet feasible. With this as background, a measurement-device-independent (MDI) QKD was proposed \cite{Lo12}. The MDI QKD is secure under the assumption that the source is ideal; that is, the source is exactly in the prescribed quantum state. The MDI QKD with uncharacterized quantum bits (qubits) adapts an assumption that the source is within a two-dimensional subspace \cite{Yin13,Yin14}. (The latter assumption is weaker than the former because the prescribed states are already two-dimensional. In the former case, we need characterization process to estimate how close the source is to the prescribed one.) The assumption of two-dimensionality is ``not too stringent in many practical and BB84 systems. For example, in phase-encoding systems, it is reasonable to assume that the encoding states are in two-dimensional subspaces...\cite{Yin14}".
The MDI QKD with un-characterized qubits was recently improved to get higher key rates and simpler derivations \cite{Hwa17}. On the other hand, a protocol that dispenses with the shared reference frame, the reference-frame-independent (RFI) QKD was proposed \cite{Lai10}. It was observed that RFI QKD is also useful for a channel with uncontrolled unitary rotation along an axis \cite{Yin14B}. The security of the RFI MDI QKD was also assured under the assumption of ideal source \cite{Yin14B}. However, the security of the RFI MDI QKD with uncharacterized qubits not yet been shown.

In this paper, we present the security of the RFI MDI QKD with uncharacterized qubits. In Section II, we briefly review the (improved) MDI QKD with uncharacterized qubits. In Section III, we show the security of the RFI MDI QKD with uncharacterized qubits. In Section IV, we discuss our results and present conclusions.
%%%%%%%%%%%%%%%%%%%%%%%%%%%%%%%%%%%%%%%
\section{MDI QKD with uncharacterized qubits}
%%%%%%%%%%%%%%%%%%%%%%%%%%%%%%%%%%%%%%%
For the protocol \cite{Yin14}, each user prepares two encoding states. Let the states prepared by Alice and Bob be denoted by $|\varphi_m\rangle$ and $|\varphi_n^{\prime}\rangle$, respectively, where $m,n=0,1$. Here, nothing is supposed about the encoding states, namely, they are uncharacterized. Each user also prepares a checking state, {\it which is assumed to be a superposition of the encoding states}. Alice's and Bob's checking states are, respectively,
 %%%%%%%%%%%%%%%%%%%%%%%%%%%%%%%%%%%%%%%%%
\begin{eqnarray}
|\varphi_2\rangle= c_0|\varphi_0\rangle + c_1 e^{i\theta}|\varphi_1\rangle,
 \nonumber
 \\
|\varphi_2^{\prime}\rangle= c_0^{\prime}|\varphi_0^{\prime}\rangle + c_1^{\prime} e^{i\theta^{\prime}}|\varphi_1^{\prime}\rangle.
\label{1}
\end{eqnarray}
%%%%%%%%%%%%%%%%%%%%%%%%%%%%%%%%%%%%%%%%%(1)
Here, $c_m$ and $c_n^{\prime}$ are non-negative numbers and, $\theta$ and $\theta^{\prime}$ are real. The protocol is as follows:

(1) Alice generates a random number $i$ where $i=0,1,2$. She sends a state $|\varphi_i\rangle$ to Charlie. Here, Charlie can be anyone, so Charlie can be either Eve (eavesdropper) or the users themselves. (2) Bob independently generates a random number $j$ where $j=0,1,2$. He also sends a state $|\varphi_{j}^{\prime}\rangle$ to Charlie. (3) Charlie performs a measurement on the states $|\varphi_i\rangle$ and $|\varphi_{j}^{\prime}\rangle$. The measurement can be any one that finally gives two outcomes $0$ and $1$. Charlie announces the outcome.
(4) When the outcome is $0$, the users discard the data. Otherwise, they keep the data. By sacrificing some of the data for public discussion, the users estimate, $p_{ij}^1$, the conditional probability to get outcome $1$ for each $i,j$.
(5) Measurement data with both $i$ and $j$ less than $2$ become the raw key. Other data with either $i$ or $j$ as $2$ are used for checking purposes. The users do postprocessing to get the final key.

Now let us consider Charlie's measurement on the states $|\varphi_i\rangle$ and $|\varphi_{j}^{\prime}\rangle$. In the most general collective attack, Eve attaches an ancilla $|e\rangle$ to the states and then applies a unitary operation to them \cite{Yin14}:
 %%%%%%%%%%%%%%%%%%%%%%%%%%%%%%%%%%%%%%%%%
\begin{eqnarray}
 && U_{\mbox{Eve}}|\varphi_i\rangle |\varphi_j^{\prime}\rangle |e\rangle |0\rangle_M  \nonumber\\
 && = \sqrt{p^0_{ij}}\hspace{1mm} |\Gamma_{ij}^0\rangle |0\rangle_M+  \sqrt{p^1_{ij}}\hspace{1mm} |\Gamma_{ij}^1\rangle |1\rangle_M.
\label{2}
\end{eqnarray}
%%%%%%%%%%%%%%%%%%%%%%%%%%%%%%%%%%%%%%%%%(2)
Eve gets the outcome by measuring the quantum state indexed by $M$ in the basis of $|0\rangle$ and $|1\rangle$. Consider only the data with the measurement outcome $1$ is sufficient because those with $0$ are discarded. For convenience, let us omit $1$ from now on; $|\Gamma_{ij}^1\rangle \equiv |\Gamma_{ij}\rangle$ and $p_{ij}^1\equiv p_{ij}$. Now we can see that Eqs. (\ref{1}) and (\ref{2}) give the constraints
%%%%%%%%%%%%%%%%%%%%%%%%%%%%%%%%%%
\begin{eqnarray}
\sqrt{p_{2n}}\hspace{1mm} |\Gamma_{2n}\rangle &=& \sqrt{p_{0n}}\hspace{1mm} c_0|\Gamma_{0n} \rangle+ \sqrt{p_{1n}}\hspace{1mm} c_1 e^{i\theta} |\Gamma_{1n} \rangle,
\label{3}
\\
\sqrt{p_{m2}}\hspace{1mm} |\Gamma_{m2}\rangle &=& \sqrt{p_{m1}}\hspace{1mm} c_0^{\prime}|\Gamma_{m0} \rangle+ \sqrt{p_{m1}}\hspace{1mm} c_1^{\prime} e^{i\theta^{\prime}} |\Gamma_{m1} \rangle,
\label{4}
\\
\sqrt{p_{22}} \hspace{1mm} |\Gamma_{22}\rangle &=& \sum_{m,n} \sqrt{p_{mn}} \hspace{1mm} c_m c_n^{\prime} e^{i\theta_m} e^{i\theta_n^{\prime}}|\Gamma_{mn}\rangle,
\label{5}
\end{eqnarray}
%%%%%%%%%%%%%%%%%%%%%%%%%%%%%%%%%%(3-5)
where $\theta_0=\theta_0^{\prime}=0$, $\theta_1=\theta$, and
$\theta_1^{\prime}=\theta^{\prime}$. The constraints (\ref{3})-(\ref{5}) play key roles in the security analysis.

For theoretical purposes, we consider an equivalent entanglement distillation protocol \cite{Sho00,Nie00}. In the hypothetical protocol, Alice and Bob prepare entangled states $(1/\sqrt{2})(|0\rangle_{A_1} |\varphi_0\rangle_{A_2} +|1\rangle_{A_1} |\varphi_1\rangle_{A_2})$ and  $(1/\sqrt{2})(|0\rangle_{B_1} |\varphi_0^{\prime}\rangle_{B_2} +|1\rangle_{B_1} |\varphi_1^{\prime}\rangle_{B_2})$, respectively. Alice and Bob send quantum states indexed by $A_2$ and $B_2$ to Charlie, respectively. According to Charlie's announcement about when the measurement outcome is $1$, the users postselect their qubits indexed by $A_1$ and $B_1$, respectively. The postselected state is given by \cite{Yin14}
%%%%%%%%%%%%%%%%%%%%%%%%%%%%%%%%%%%%%%%%
\begin{eqnarray}
&&\rho= \frac{1}{p_{00}+p_{11}+p_{01}+p_{10}} \cdot \nonumber\\
&&\sum_q \bold{P}[\sqrt{p_{00}} \gamma_{00}^q |0\rangle_{A_1}|0\rangle_{B_1}+ \sqrt{p_{11}} \gamma_{11}^q |1\rangle_{A_1}|1\rangle_{B_1} \nonumber\\
&&+\sqrt{p_{01}} \gamma_{01}^q |0\rangle_{A_1}|1\rangle_{B_1}+ \sqrt{p_{10}} \gamma_{10}^q |1\rangle_{A_1}|0\rangle_{B_1}].
\label{6}
\end{eqnarray}
%%%%%%%%%%%%%%%%%%%%%%%%%%%%%%%%%%%%%%%%(6)
Here $\bold{P}[x]\equiv |x\rangle \langle x|$, $|\Gamma_{mn}\rangle \equiv \sum_{q} \gamma_{mn}^q |q\rangle$, where $|q\rangle$ is a set of orthonormal states. From normalization, $\sum_q |\gamma_{mn}^q|^2=1$. We consider four Bell states $|\varphi^{\pm \alpha} \rangle= (1/\sqrt{2})(|00\rangle \pm e^{i(\alpha_A+ \alpha_B)}|11\rangle)$ and $|\psi^{\pm \alpha} \rangle= (1/\sqrt{2})(|01\rangle \pm e^{i(\alpha_A- \alpha_B)}|10\rangle)$, which are obtained by rotating each qubit by Pauli operator $\sigma_z$ with amounts $\alpha_A$ and $\alpha_B$, respectively. The final key rate is given by $R= 1-H(e_b)-H(e_p)$, where $H(x)= -x\log_2 x- (1-x)\log_2 (1-x)$ is the binary Shannon entropy, and $e_b$ and $e_p$ are the bit and the phase error rates, respectively.
The bit error rate is given by
 %%%%%%%%%%%%%%%%%%%%%%%%%%%%%%%%%%%%%%%%%
\begin{eqnarray}
 e_b&=&\langle \psi^{+\alpha}| \rho |\psi^{+\alpha} \rangle +
     \langle \psi^{-\alpha}| \rho |\psi^{-\alpha} \rangle
     \nonumber\\
    &=&\frac{p_{01}+p_{10}}{p_{00}+p_{11}+p_{01}+p_{10}}
\label{7}
\end{eqnarray}
%%%%%%%%%%%%%%%%%%%%%%%%%%%%%%%%%%%%%%%%%%%(7)
for arbitrary $\alpha_A$ and $\alpha_B$.
The phase error rate is given by
%%%%%%%%%%%%%%%%%%%%%%%%%%%%%%%%%%%%%%%%%
\begin{eqnarray}
 e_p&=&\langle \varphi^{-\alpha}|\rho|\varphi^{-\alpha}\rangle +
       \langle \psi^{-\alpha}| \rho |\psi^{-\alpha} \rangle
       \nonumber\\
    &\leq& e_b+ \frac{\sum_q|\sqrt{p_{00}}\gamma_{00}^q- e^{i(\alpha_A+ \alpha_B)} \sqrt{p_{11}} \gamma_{11}^q|^2}{2(p_{00}+p_{11}+p_{01}+p_{10})}.
\label{8}
\end{eqnarray}
%%%%%%%%%%%%%%%%%%%%%%%%%%%%%%%%%%%%%%%%%%%%%(8)
Let us consider the quantity to be bounded. We get
%%%%%%%%%%%%%%%%%%%%%%%%%%%%%%%%%%%%%%%%%
\begin{eqnarray}
&&\Delta = \sum_q|\sqrt{p_{00}}\gamma_{00}^q- e^{i(\alpha_A+ \alpha_B)} \sqrt{p_{11}} \gamma_{11}^q|^2 \nonumber\\
&=& p_{00}+p_{11}-2\sqrt{p_{00}}\sqrt{p_{11}}\hspace{1mm}
\mbox{Re}\hspace{0.5mm}[e^{i(\alpha_A+ \alpha_B)}\langle\Gamma_{00}|\Gamma_{11}\rangle ], \nonumber\\
\label{9}
\end{eqnarray}
%%%%%%%%%%%%%%%%%%%%%%%%%%%%%%%%%%%%%%%%%%%(9)
where Re$\hspace{0.5mm}[z]$ is the real part of a complex number $z$ \cite{Hwa17}. From constraint (\ref{5}), we get
%%%%%%%%%%%%%%%%%%%%%%%%%%%%%%%%%%%%%%%%%
\begin{eqnarray}
&&\mbox{Re}\hspace{0.5mm}[e^{i(\theta+\theta^{\prime})}
\langle\Gamma_{00}|\Gamma_{11}\rangle] \leq
\nonumber\\
&&
\frac{|\sqrt{p_{22}}+\sqrt{p_{01}} c_0 c_1^{\prime}
+\sqrt{p_{10}} c_1 c_0^{\prime}|^2-p_{00} c_0^2 {c_0^{\prime}}^2- p_{11} c_1^2 {c_1^{\prime}}^2}{2\sqrt{p_{00}}\sqrt{p_{11}} c_0 c_0^{\prime} c_1 c_1^{\prime}}
\nonumber\\
\label{10}
\end{eqnarray}
%%%%%%%%%%%%%%%%%%%%%%%%%%%%%%%%%%%%%%%%%%%(10)
by $||\hspace{0.5mm}|A\rangle +|B\rangle|| \leq ||\hspace{0.5mm}|A\rangle||+||\hspace{0.5mm}|B\rangle ||$ where $|\hspace{0.5mm}A\rangle$, $|\hspace{0.5mm}B\rangle$ are arbitrary states and $||\hspace{0.5mm}|A\rangle||$ is the norm of $|A\rangle$.
We numerically upper bound $\mbox{Re}\hspace{0.5mm}[e^{i(\theta+\theta^{\prime})} \langle\Gamma_{00}|\Gamma_{11}\rangle]$ by using the inequality (\ref{10}) with the constraints (\ref{1}),(\ref{3}), and (\ref{4}) (Also refer to Eqs. (9),(10), and (14) in Ref. \cite{Hwa17}.)

Consider a case when the upper bound is negative, that is, $\mbox{Re}\hspace{0.5mm}[e^{i(\theta+\theta^{\prime})} \langle\Gamma_{00}|\Gamma_{11}\rangle]\leq -|\Omega|$; then we can see that $|\langle\Gamma_{00}|\Gamma_{11}\rangle|$ is lower bounded by the absolute value of the upper bound,
$|\langle\Gamma_{00}|\Gamma_{11}\rangle| \geq |\Omega|$.
Now it is easy to see that we can always choose $\alpha_A+ \alpha_B$ such that
%%%%%%%%%%%%%%%%%%%%%%%%%%%%%%%%%%%%%%%%%
\begin{equation}
\Delta \leq p_{00}+ p_{11}- 2 \sqrt{p_{00}} \sqrt{p_{11}} \hspace{0.5mm} |\Omega|.
\label{11}
\end{equation}
%%%%%%%%%%%%%%%%%%%%%%%%%%%%%%%%%%%%%%%%%%%(11)
Now we can get the upper bound for the phase error rate by inequalities (\ref{8}) and (\ref{11}), from which we can get the final key rate \cite{Hwa17}.
%%%%%%%%%%%%%%%%%%%%%%%%%%%%%%%%%%%%%%%
\section{RFI MDI QKD with uncharacterized qubits}
%%%%%%%%%%%%%%%%%%%%%%%%%%%%%%%%%%%%%%%
Consider a channel affected by an uncontrolled unitary rotation along an axis. Unless actively adjusted, normally the rotation increases bit or phase errors such that no key remains. However, even without active adjustment, RFI QKD's enable key generation in this situation \cite{Lai10,Yin14B}. Here we provide the security of the RFI MDI QKD with uncharacterized qubits:
First we introduce more checking states and then we get more bounds for the real part of the inner product  $\langle\Gamma_{00}|\Gamma_{11}\rangle$ (multiplied by some phase). Using the bounds, we lower bound absolute value of the inner product. Then we can get final key rate from the absolute value of the inner product, as in Section II.

The RFI MDI QKD with uncharacterized qubits is almost the same as the MDI QKD with uncharacterized qubits described in Section II. The only difference is that, Alice and Bob each generate one more checking-state
$|\varphi_3\rangle= \tilde{c}_0|\varphi_0\rangle + \tilde{c}_1 e^{i\tilde{\theta}}|\varphi_1\rangle$ and
$|\varphi_3^{\prime}\rangle= \tilde{c}_0^{\prime}|\varphi_0^{\prime}\rangle + \tilde{c}_1^{\prime} e^{i\tilde{\theta}^{\prime}}|\varphi_1^{\prime}\rangle$, respectively, so $i,j=0,1,2,3$ here. (In Ref. \cite{Yin14}, the case of two checking states for each user is also considered, but actually only a pair of checking states is used.) Thus four combinations of checking states exist: $(2,2)$, $(2,3)$, $(3,2)$, and $(3,3)$. For each combination, using an inequality corresponding to the inequality (\ref{10}),
we independently upper bound real part of inner product $\langle\Gamma_{00}|\Gamma_{11}\rangle$ multiplied by some phase. For example, for the $(2,3)$ pair of checking states, we get the upper bound by using
%%%%%%%%%%%%%%%%%%%%%%%%%%%%%%%%%%%%%%%%%
\begin{eqnarray}
&&\mbox{Re}\hspace{0.5mm}[e^{i(\theta+\tilde{\theta}^{\prime})}
\langle\Gamma_{00}|\Gamma_{11}\rangle] \leq
\nonumber\\
&&
\frac{|\sqrt{p_{23}}+\sqrt{p_{01}} c_0 \tilde{c}_1^{\prime}
+\sqrt{p_{10}} c_1 \tilde{c}_0^{\prime}|^2-p_{00} c_0^2 \tilde{c_0^{\prime}}^2- p_{11} c_1^2 \tilde{c_1^{\prime}}^2}{2\sqrt{p_{00}}\sqrt{p_{11}} c_0 \tilde{c}_0^{\prime} c_1 \tilde{c}_1^{\prime}}.
\nonumber\\
\label{12}
\end{eqnarray}
%%%%%%%%%%%%%%%%%%%%%%%%%%%%%%%%%%%%%%%%%%%(12)
Consequently we get four upper bounds: one each for $\mbox{Re}\hspace{0.5mm}[e^{i(\theta+\theta^{\prime})}
\langle\Gamma_{00}|\Gamma_{11}\rangle]$, $\mbox{Re}\hspace{0.5mm}[e^{i(\theta+\tilde{\theta}^{\prime})}
\langle\Gamma_{00}|\Gamma_{11}\rangle]$
$\mbox{Re}\hspace{0.5mm}[e^{i(\tilde{\theta}+\theta^{\prime})}
\langle\Gamma_{00}|\Gamma_{11}\rangle]$, and
$\mbox{Re}\hspace{0.5mm}[e^{i(\tilde{\theta}+\tilde{\theta}^{\prime})}
\langle\Gamma_{00}|\Gamma_{11}\rangle]$.

From the constraint (\ref{5}), we can also get lower bounds for the four quantities. (Also refer to Eq. (14) in Ref. \cite{Hwa17}.) For example, for the $(2,3)$ pair,
%%%%%%%%%%%%%%%%%%%%%%%%%%%%%%%%%%%%%%%%%
\begin{eqnarray}
&&\mbox{Re}\hspace{0.5mm}[e^{i(\theta+\tilde{\theta}^{\prime})}
\langle\Gamma_{00}|\Gamma_{11}\rangle] \geq
\nonumber\\
&&
\frac{|\sqrt{p_{23}}-\sqrt{p_{01}} c_0 \tilde{c}_1^{\prime}
-\sqrt{p_{10}} c_1 \tilde{c}_0^{\prime}|^2-p_{00} c_0^2 \tilde{c_0^{\prime}}^2- p_{11} c_1^2 \tilde{c_1^{\prime}}^2}{2\sqrt{p_{00}}\sqrt{p_{11}} c_0 \tilde{c}_0^{\prime} c_1 \tilde{c}_1^{\prime}}.
\nonumber\\
\label{13}
\end{eqnarray}
%%%%%%%%%%%%%%%%%%%%%%%%%%%%%%%%%%%%%%%%%%%(13)
Here we used $||\hspace{0.5mm}|A\rangle +|B\rangle+ |C\rangle|| \geq ||\hspace{0.5mm}|A\rangle||-||\hspace{0.5mm}|B\rangle ||-||\hspace{0.5mm}|C\rangle ||$. Note that the four quantities are now between the lower and the upper bounds. For example, the quantity $\mbox{Re}\hspace{0.5mm}[e^{i(\theta+\tilde{\theta}^{\prime})}
\langle\Gamma_{00}|\Gamma_{11}\rangle]$ is between the  bounds in Eqs. (\ref{12}) and (\ref{13}).

Using the bounds for the four quantities, we can lower bound the absolute value of the inner product. Let us suppose that the channel undergoes depolarization and a (unitary)  rotation along $z$ with angle $\Theta$. We denote a qubit in the $x$-$y$ plane in the Bloch sphere \cite{Nie00}, $(1/\sqrt{2})(|0\rangle+ e^{i \phi} |1\rangle)$, by $|\Phi= \phi\rangle$. For examples, $|\Phi= 0 \rangle$ denotes $(1/\sqrt{2})(|0\rangle+ |1\rangle)$ and $|\Phi=  \pi/2 \rangle$ denotes $(1/\sqrt{2})(|0\rangle+ i|1\rangle)$. Now let us suppose that
$|\varphi_0\rangle= |0\rangle$, $|\varphi_0^{\prime}\rangle= |1\rangle$, $|\varphi_1\rangle= |1\rangle$, $|\varphi_1^{\prime}\rangle= |0\rangle$, $|\varphi_2\rangle= |\Phi= 0\rangle$, $|\varphi_2^{\prime}\rangle= |\Phi= \pi\rangle$, and $|\varphi_3\rangle= |\pi/2\rangle$ $|\varphi_3^{\prime}\rangle= |\Phi= |3\pi/2\rangle$, and that the measurement element for outcome $1$ is $|\psi^-\rangle= (1/\sqrt{2})(|0 \rangle |1\rangle- |1\rangle |0\rangle)$.
Then the probabilities corresponding to the bit error rate $e_b$ are $p_{00}=p_{11}=(1/2)(1-e_b), p_{10}=p_{01}=(1/2)e_b, p_{20}=p_{21}=p_{02}=p_{12}=p_{30}=p_{31}=p_{03}=p_{13}=1/4$, and %%%%%%%%%%%%%%%%%%%%%%%%%%%%%%%%%%%%%%%%%%%%%%%
\begin{eqnarray}
p_{22}&=&(1-e_b)\frac{1}{4}(1+\cos\Theta)+ e_b \frac{1}{2},
\nonumber\\
p_{23}&=&(1-e_b)\frac{1}{4}(1+\cos[\frac{\pi}{2}+\Theta])+ e_b \frac{1}{2},
\nonumber\\
p_{32}&=&(1-e_b)\frac{1}{4}(1+\cos[\frac{\pi}{2}-\Theta])+ e_b \frac{1}{2},
\nonumber\\
p_{33}&=&(1-e_b)\frac{1}{4}(1+\cos\Theta)+ e_b \frac{1}{2}.
\label{14}
\end{eqnarray}
%%%%%%%%%%%%%%%%%%%%%%%%%%%%%%%%%%%%%%%%%%%(14)

First let us consider an ideal case of zero depolarization. Then we have $p_{00}=p_{11}=1/2,  p_{10}=p_{01}=0, p_{20}=p_{21}=p_{02}=p_{12}=p_{30}=p_{31}=p_{03}=p_{13}=1/4$, $p_{22}= p_{33}= (1/4)(1+\cos\Theta)$, $p_{23}=\frac{1}{4}(1+\cos[\frac{\pi}{2}+\Theta])
$, and $ p_{32}=(1/4)(1+\cos[\frac{\pi}{2}-\Theta])$. Now one can easily see that
%%%%%%%%%%%%%%%%%%%%%%%%%%%%%%%%%%%%%%%%%%%%%%%
\begin{eqnarray}
\mbox{Re}\hspace{0.5mm}[e^{i(\theta+\theta^{\prime})}
\langle\Gamma_{00}|\Gamma_{11}\rangle]&=&\cos\Theta,
\nonumber\\
\mbox{Re}\hspace{0.5mm}[e^{i(\theta+\tilde{\theta}^{\prime})}
\langle\Gamma_{00}|\Gamma_{11}\rangle]&=&\cos[\frac{\pi}{2}+\Theta],
\nonumber\\
\mbox{Re}\hspace{0.5mm}[e^{i(\tilde{\theta}+\theta^{\prime})}
\langle\Gamma_{00}|\Gamma_{11}\rangle]&=&\cos[\frac{\pi}{2}-\Theta],
\nonumber\\
\mbox{Re}\hspace{0.5mm}[e^{i(\tilde{\theta}+\tilde{\theta}^{\prime})}
\langle\Gamma_{00}|\Gamma_{11}\rangle]&=& \cos\Theta;
\label{15}
\end{eqnarray}
%%%%%%%%%%%%%%%%%%%%%%%%%%%%%%%%%%%%%%%%%%%(15)
the normalization for Eq. (\ref{1}) and the constraints (\ref{3}) and (\ref{4}), combined with the probabilities, give $c_0= c_1= c_0{\prime}= c_1{\prime}= (1/\sqrt{2})$. Then combined with the inequalities (\ref{12}) and (\ref{13}), and the probabilities, we can get the second expression in Eq. (\ref{15}). The others can be obtained in the same way. Let us consider the case $\Theta= \pi/4$, (which corresponds to the maximal violation of the Bell inequality \cite{Nie00}), or $\mbox{Re}\hspace{0.5mm}[e^{i(\theta+\theta^{\prime})}
\langle\Gamma_{00}|\Gamma_{11}\rangle]= 1/\sqrt{2},
\mbox{Re}\hspace{0.5mm}[e^{i(\theta+\tilde{\theta}^{\prime})}
\langle\Gamma_{00}|\Gamma_{11}\rangle]= -1/\sqrt{2},
\mbox{Re}\hspace{0.5mm}[e^{i(\tilde{\theta}+\theta^{\prime})}
\langle\Gamma_{00}|\Gamma_{11}\rangle]= 1/\sqrt{2}$, and
$\mbox{Re}\hspace{0.5mm}[e^{i(\tilde{\theta}+ \tilde{\theta}^{\prime})}\langle\Gamma_{00}|\Gamma_{11}\rangle]= 1/\sqrt{2}$. If we set $\langle\Gamma_{00}|\Gamma_{11}\rangle= |\langle\Gamma_{00}|\Gamma_{11}\rangle|e^{i \delta}= re^{i \delta}$, $\delta+ \theta+ \theta^{\prime}= A$, $\tilde{\theta}^{\prime}- \theta^{\prime}= B$, and $\tilde{\theta}- \theta= C$, we obtain $\mbox{Re}\hspace{0.5mm}[re^{iA}]= 1/\sqrt{2},
\mbox{Re}\hspace{0.5mm}[re^{i(A+B)}]= -1/\sqrt{2},
\mbox{Re}\hspace{0.5mm}[re^{i(A+C)}]= 1/\sqrt{2}$, and
$\mbox{Re}\hspace{0.5mm}[re^{i(A+B+C)}]= 1/\sqrt{2}$.
We can see that these four equalities can be satisfied only by that the inner product $r=1$ with $A=\pm\pi/4, B=\pm \pi/2, C=\mp \pi/2$, by either geometric intuition or numerical method. From the lower bound on the size of the inner product $r$, we can calculate the final key rate; the lower bound on the size of the inner product gives the difference between the phase error and the bit error rates, $[p_{00}+ p_{11}- 2 \sqrt{p_{00}} \sqrt{p_{11}} \hspace{0.5mm} |\Omega|]/[2(p_{00}+p_{11}+p_{01}+p_{10})]$, where $|\Omega|$ is the lowerbound on the inner product size $r$. The final key rate is given by $R= 1-H(e_b)-H(e_p)$. Thus the size of the inner product $r=1$ means that the final key rate is one.

Now let us consider the realistic non-zero depolarization case. The four quantities
$\mbox{Re}\hspace{0.5mm}[e^{iA}
\langle\Gamma_{00}|\Gamma_{11}\rangle]$, $\mbox{Re}\hspace{0.5mm}[e^{i(A+B)}
\langle\Gamma_{00}|\Gamma_{11}\rangle]$
$\mbox{Re}\hspace{0.5mm}[e^{i(A+C)}
\langle\Gamma_{00}|\Gamma_{11}\rangle]$, and
$\mbox{Re}\hspace{0.5mm}[e^{i(A+B+C)}
\langle\Gamma_{00}|\Gamma_{11}\rangle]$
are between the upper and the lower bounds, as we have seen.
The upper and the lower bounds can be numerically obtained; numerical optimization of the inequalities (\ref{12}) and (\ref{13}) with the probabilities just above Eq. (\ref{14}) and the constraints (\ref{1}),(\ref{3}), and (\ref{4}), gives the upper and the lower bounds for the second quantity $\mbox{Re}\hspace{0.5mm}[e^{i(A+B)}
\langle\Gamma_{00}|\Gamma_{11}\rangle]$. Bounds for other three quantities can be obtained in the same way.
Then we can get the lower bound on the size of the inner product by using a numerical method and the bounds on the four quantities.
Then we obtain the final key rate, which is shown in Fig 1.
%%%%%%%%%%%%%%%%%%%%%%%%%%%%%%%%%%%%%%%%%
\begin{figure}
\includegraphics[width=8cm]{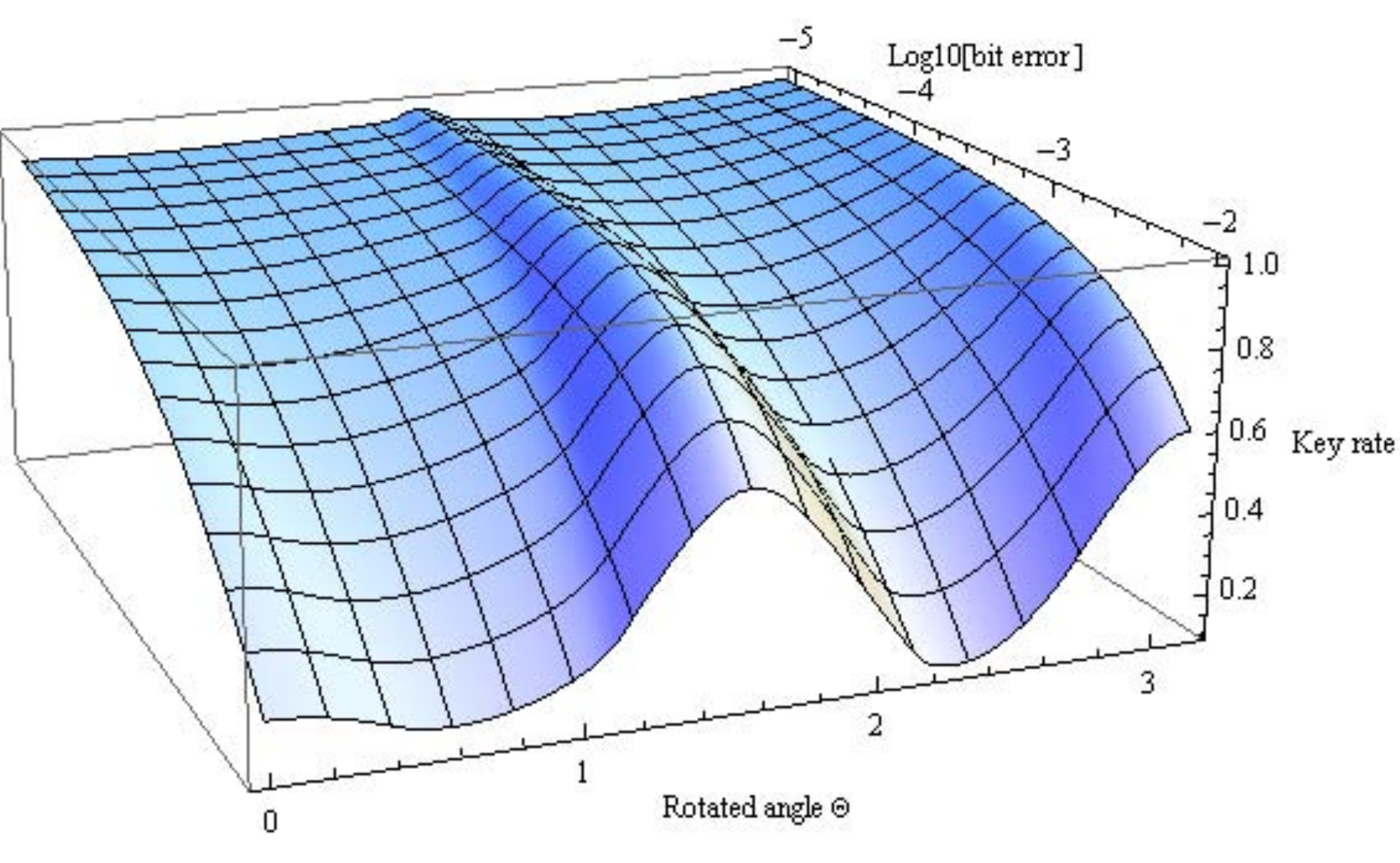}
\caption{Final key rates for various angles and bit error rates. The final key rates for angles from $\pi$ to $2\pi$ are the same as those for angles from $0$ to $\pi$ which is shown.}
\label{Fig-1}
\end{figure}
%%%%%%%%%%%%%%%%%%%%%%%%%%%%%%%%%%%%%%%%(Fig. 1)
Here we can see that when the bit error rate $e_b$ is less than around $1\%$, the final key rate is non-zero regardless of rotation angle. Although the bit error rate's threshold is quite low, there is an advantage; active control of the rotation is not necessary. In particular, the protocol is very useful for an implementation, such as the which-path qubit implementation \cite{Yin14B}, with a channel that has a very low bit error rate but suffers uncontrolled unitary rotation. Here the unitary rotation may be uncontrolled but should be slow enough such that amount of accumulated data for a certain fixed angle is large enough for post-processing to extract the key. One might say that even in the non-RFI MDI QKD keys may be distributed if the rotation is slow. However, in the non-RFI case, active angle adjustment in the hardware of the protocol is necessary while in our protocol no physical adjustment is necessary. Even when the rotation is not slow, key generation is still possible \cite{Pra17} but the protocol is not MDI; thus the problem of imperfect detectors remains \cite{Mak06, Qi07, Fun07}.
%%%%%%%%%%%%%%%%%%%%%%%%%%%%%%%%%%%%%%%
\section{Discussion and Conclusion}
%%%%%%%%%%%%%%%%%%%%%%%%%%%%%%%%%%%%%%%
For angles $0$, $\pi/2$, and $\pi$, the final key rates are higher than those for other angles. This is because in these cases, at least one of the probabilities $p_{22}$, $p_{23}$, $p_{32}$, or $p_{33}$ has either the maximum value $1/2$ or minimum value $0$; In the MDI QKD with uncharacterized qubits, when the probabilities for checking states is either maximized or minimized, the gap between the phase and the bit error rates become small; thus more keys are generated. As is also observed in Fig 1, the minimum case, $\Theta= \pi$, is more efficient than the maximum case, $\Theta= 0$. Thus usually the minimum case is adapted in the (non-RFI) MDI QKD with uncharacterized qubits \cite{Yin14,Hwa17}.

To summarize, we proposed a RFI MDI QKD with uncharacterized qubits. We showed that key cound be generated for a bit error rate up to around $1\%$ for a channel of uncontrolled, (but slow) unitary rotation in addition to depolarization. The protocol can also be useful for the implementation with that channel.
\section*{Acknowledgement}
We are grateful to Dr. Hongyi Su for assisting with the numerical works. This study was supported by Institute for Information and Communications Technology Promotion (IITP) grant funded by the Korea Government (MSIP) (No. R0190-18-2028, Practical and Secure Quantum Key Distribution).

%%%%%%%%%%%%%%%%%%%%%%%%%%%%%%%%%%%%%%%%%%%%%%%%%%%%%%%%%%%%
%%%%%%%%%%%%%%%%%%%%%%%%%%%%%%%%%%%%%%%%%%%%%%%%%%%%

\end{document}